\documentclass[preprint]{aastex}
\usepackage{emulateapj5,epsfig}

\newcommand{\kms}{$\rm km \; \rm s^{-1}$}  
\def\deg{$^\circ$\hspace{-1.7mm}.\hspace{0.3mm}}  
\def\as{\arcsec\hspace{-1.7mm}.\hspace{0.3mm}}

\received{---------}  
\revised{---------}  
\accepted{---------}  
\journalid{}{}  
\articleid{}{}  
  
\slugcomment{Accepted for publication in the Astrophysical Journal 2001, 
v.~555}  
  
\lefthead{Patat et al.}  
\righthead{The metamorphosis of SN~1998bw}  
  
\begin{document}  
  
\title{The metamorphosis of SN~1998bw$^\ddagger$}  
  
\author{Ferdinando Patat\altaffilmark{1}, Enrico Cappellaro\altaffilmark{2},  
John Danziger\altaffilmark{3}, Paolo A. Mazzali\altaffilmark{3,4}, 
Jesper Sollerman\altaffilmark{1,5}, 
Thomas Augusteijn\altaffilmark{6}, James Brewer\altaffilmark{6},    
Vanessa Doublier\altaffilmark{6}, Jean Fran\c{c}ois Gonzalez\altaffilmark{6,7},   
Olivier Hainaut\altaffilmark{6}, Chris Lidman\altaffilmark{6},   
Bruno Leibundgut\altaffilmark{1}, Ken'ichi Nomoto\altaffilmark{4}, 
Takayoshi Nakamura\altaffilmark{4}, 
Jason Spyromilio\altaffilmark{1}, Luca Rizzi\altaffilmark{2},   
Massimo Turatto\altaffilmark{2}, Jeremy Walsh\altaffilmark{1}, 
Titus J. Galama\altaffilmark{8},   
Jan van Paradijs\altaffilmark{8,9}$^{,\star}$,   
Chryssa Kouveliotou\altaffilmark{10},   
Paul M. Vreeswijk\altaffilmark{8}, Filippo Frontera\altaffilmark{11},   
Nicola Masetti\altaffilmark{11}, Eliana Palazzi\altaffilmark{11},   
Elena Pian\altaffilmark{11}}  
  
\vspace{5mm}  
$\ddagger$ Based on observations collected at ESO--La Silla.  
  
$\star$ Deceased on November 2, 1999.  
  
\altaffiltext{1}{European Southern Observatory, Karl Schwarzschild Str.~2,   
D-85748 Garching b. M\"unchen, Germany; E-mail: fpatat@eso.org}  
  
\altaffiltext{2}{Osservatorio Astronomico di Padova, v. Osservatorio 5,   
I-35122 Padova, Italy} 
 
\altaffiltext{3}{Osservatorio Astronomico di Trieste, v. G. B.   
Tiepolo 11, I-34131 Trieste, Italy}  
  
\altaffiltext{4}{Research Centre for the Early Universe and Department of   
Astronomy, University of Tokyo, Bunkyo-ku, Tokyo 113-0033, Japan}  
  
\altaffiltext{5}{Stockholm Observatory, S-133 36, Saltsj\"obaden, Sweden}  
 
\altaffiltext{6}{European Southern Observatory, A. de Cordova 3107,   
Casilla 19001 Santiago, Chile}  
  
\altaffiltext{7}{Centre de Recherche Astronomique de Lyon, (CNRS-UMR 5574)  
Ecole Normale Sup\'erieure 46 all\'ee d'Italie  
F-69364 Lyon Cedex 07, France}  
  
\altaffiltext{8}{Astronomical Institute Anton Pannekoek \ CHEAF,   
Kruislaan 403, 1098 SJ, Amsterdam, The Netherlands}  
  
\altaffiltext{9}{Department of Physics, University of Alabama in Huntsville,  
Huntsville, AL35899, USA}  
  
\altaffiltext{10}{NASA Marshall Space Flight Center, SD-40, Huntsville,   
Alabama 35812, USA}  
  
\altaffiltext{11}{Istituto Tecnologie e Studio Radiazioni Extraterrestri,   
CNR, Bologna, Italy}

\setcounter{footnote}{0} 
  
\begin{abstract}  
  
We present and discuss the photometric and spectroscopic evolution of the   
peculiar SN~1998bw, associated with GRB~980425, through an   
analysis of optical and near IR data collected at ESO-La Silla.  
    
The spectroscopic data, spanning the period from day $-9$ to day +376 
(relative to $B$ maximum), have shown that this SN was unprecedented, 
although somewhat similar to SN~1997ef.  Maximum expansion velocities 
as high as $3\times10^4$ \kms\/ to some extent mask its resemblance to 
other Type Ic SNe. At intermediate 
phases, between photospheric and fully nebular, the expansion velocities 
($\sim 10^{4}$ \kms) remained exceptionally high compared to those of 
other recorded core-collapse SNe at a similar phase. 
The mild linear polarization detected at early epochs 
suggests the presence of asymmetry in the emitting material. The degree of  
asymmetry, however, cannot be decoded from these measurements alone. 

The He~I 1.083~$\mu m$ and 2.058~$\mu m$ lines are identified and He is suggested to 
lie in an outer region of the envelope. The temporal behavior of the fluxes and 
profiles of emission lines of Mg~I]~4571~\AA, [O~I]~6300,6364~\AA\/ and  
a feature ascribed to Fe are traced to stimulate future modeling work. 

The uniqueness of SN~1998bw became less obvious once it entered the 
fully nebular phase (after one year) when it was very similar to 
other Type Ib/c-IIb objects, such as the Type Ib SN~1996N and  the Type IIb 
SN~1993J, even though SN~1998bw was 1.4 magnitudes brighter than SN~1993J  
and 3 magnitudes brighter than SN~1996N at a comparable phase. 
   
The late phase optical photometry, which extends up to 403 days after B 
maximum, shows that the SN luminosity declined exponentially but 
substantially faster than the decay rate of $^{56}$Co. The UVOIR bolometric  
light curve, constructed using all available optical data and the early $JHK$ 
photometry presented in this work, shows a slight flattening starting on about day +300.
Since no clear evidence of ejecta--wind interaction was found in the late time 
spectroscopy (see also Sollerman et al. 2000), this may be due to the contribution 
of the positrons as most $\gamma$--rays escape thermalization at this phase. A 
contribution from the superposed HII region can, however, not be excluded. 
\end{abstract}  
  
  
\section{Introduction}  

\begin{figure*}
\epsscale{0.8}
\caption{\label{fig:map} Upper panel: SN~1998bw in ESO 184-G82.  
The frame was obtained on June 17, 1999 (403 days after B maximum light) in  
the R band (5 min) with the ESO 3.6m+EFOSC2. 
Lower panel: HST+STIS image of SN~1998bw environment at about 764 
days after B maximum light (Fynbo et al. 2000).  
The SN location is marked by a small circle and the big circle centered 
on the SN position has a diameter of 1\arcsec.}
\epsscale{1.0} 
\end{figure*}

SN~1998bw was discovered by Galama et al. (1998a) in the BeppoSAX Wide Field 
Camera error box of GRB~980425 (Soffita et al. 1998, Pian et al. 1999) close 
to a spiral arm of the barred galaxy ESO~184--G82, by comparing two frames 
taken at the ESO New Technology Telescope (NTT) on Apr 28.4 and May 
1.3 UT.  Spectroscopic and photometric observations, both in the 
optical and in the near IR, started at ESO--La Silla immediately after the 
discovery, and showed that this object was profoundly different from 
all then known SNe (Lidman et al. 1998). The peculiar spectrum led to 
diverse classifications. A few days after its detection, the object 
was classified as a SN~Ib (Sadler et al. 1998) and later as a peculiar 
SN~Ic (Patat \& Piemonte 1998a, Filippenko 1998) owing to the complete 
absence of H lines, the weakness of the Si~II 6355~\AA\/ line and no clear 
He~I detection in the optical spectra.
  
Its peculiar spectroscopic appearance (Galama et al. 1998b, Iwamoto et 
al. 1998), its unusually high radio luminosity at early phases (Kulkarni 
et al. 1998) and, in particular, the probable association with GRB~980425 
through positional and temporal coincidence (Galama et al. 1998b, Pian 
et al. 1999) placed SN~1998bw at the center of discussion concerning the 
nature of Gamma Ray Bursts (Wheeler 2001).
 
Independent photometric and spectroscopic data sets have been presented by 
several authors, whose results will be discussed and compared with those 
presented here throughout the paper. Here we give only a brief account of
the main results of these previous works.

The early light curve of SN~1998bw has shown that the object was unusually
bright when compared to known SNe of Type Ib/c (M$_V\sim-$19.2+5log h$_{65}$,
Galama et al. 1998b). The broad-band photometric observations by
McKenzie \& Schaefer (1999) taken during the intermediate phases 
(47--171 days after maximum brightness) revealed that the object 
settled on an exponential decay similar to that observed 
in other Ic.  McKenzie \& Schaefer first suggested that even in this 
case the light curve was powered by the radioactive decay of $^{56}$Co 
with some leakage of $\gamma$--rays. 
Finally, the late phase light curve was covered up to 500 days from the
explosion by the observations of Sollerman et al. (2000). Their models 
achieved a fairly good reproduction of the data with the radioactive 
material well mixed in the ejecta and M($^{56}$Ni)$\sim$0.5 M$_\odot$.   

The peculiar spectroscopic behavior of SN~1998bw around maximum light
has been presented and discussed by Iwamoto et al. (1998), who identified
the main spectral features as O~I, Ca~II, Si~II and Fe~II. The estimated
expansion velocities were exceptionally high ($\sim$30,000 \kms) and this 
caused a severe line blending. The evolution during the first months 
was unusually slow compared to known Ic, with the nebular spectra still 
retaining many of the features present during the photospheric phase
(Patat et al. 2000, Stathakis et al. 2000). The late onset of the fully 
nebular phase has been interpreted as an indication for a large ejected mass 
(Stathakis et al. 2000) as it was predicted by the early light curve models 
(see below). During the intermediate phase, the emission lines were definitely 
broader than in known Type Ib/c supernovae and the simultaneous presence of 
iron--peak
and $\alpha$-elements indicated unusual relative abundances or physical
conditions in the SN ejecta (Patat \& Piemonte 1998b, Patat et al. 2000).
The late spectroscopy presented by Sollerman et al. (2000) showed that
the tentative morphological classification of SN~1998bw as a Type Ic 
event was indeed appropriate. The main features have been identified as
[O~I], Ca~II, Mg~I and Na~I~D, the latter possibly contaminated by 
He~I~5876~\AA.

Photometric and spectroscopic modeling of the early phases and the possible  
connection with GRB~980425 have been discussed by Galama et al.   
(1998b), Iwamoto et al. (1998), H\"oflich, Wheeler \& Wang (1999), Woosley, 
Eastman \& Schmidt (1999) and Pian et al. (1999).

The symmetric models of the early spectra and light curve by Iwamoto et 
al. (1998) and Woosley et al. (1999) reached a similar conclusion:
SN~1998bw was generated by an extremely energetic explosion
of a $\sim$10 M$_\odot$ C--O star, which ejected 0.5--0.7 
M$_\odot$ of $^{56}$Ni. This large amount of radioactive material,
which is one order of magnitude larger than that estimated for known 
core-collapse SNe (Patat et al. 1994, Schmidt et al. 1994), is
required to power the early light curve. 
In these models the explosion energy must be large to accelerate the 
ejected mass ($\sim$10 M$_\odot$) to the observed velocities and
to make the light curve peak in only 19 days. Actually, in spherical 
symmetry, the light curve around maximum can be reproduced using different 
combinations of explosion energy and envelope mass. This degeneracy was
resolved by Iwamoto et al. (1998) using velocity information and computing 
synthetic spectra for the various candidate models. 

An alternative scenario was proposed by H\"oflich et al. (1999), who used 
less energetic asymmetric models and a suitable combination of viewing angle 
and degree of asymmetry. The required mass of $^{56}$Ni is in this case 
0.2 M$_\odot$. This low mass of radioactive material, however, seems too low 
to explain the late emission of the supernova (Sollerman et al. 2000).

In this paper we report on the results of an extensive observational campaign 
carried out at ESO--La Silla, which covered the evolution of SN~1998bw 
from the discovery until 417 days after the Gamma Ray Burst detection. The 
paper is organized as follows. 
 
In Sec.~\ref{sec:obs} we discuss the observations and reduction techniques  
for the optical and the near IR. 
The evolution of SN~1998bw around maximum light is discussed 
in Sec.~\ref{sec:spec}, \ref{sec:polar} and \ref{sec:highres}, which deal 
with low resolution spectroscopy, polarimetry and high resolution spectroscopy 
respectively. The He~I detection in the near IR spectra is presented in 
Sec.~\ref{sec:helium} in which possible alternatives to the He~I 
identification are also investigated. The description of SN~1998bw's 
metamorphosis ends in Sec.~\ref{sec:nebular}, where we follow its evolution 
into the nebular phase. 
The late phase light curves are presented and compared with those  
of other SNe in Sec.~\ref{sec:lcurves}. This section also presents the  
UVOIR bolometric light curve, which is compared to the model of Iwamoto et
al. (1998). Finally, Sec.~\ref{sec:disc} summarizes our conclusions.

\section{\label{sec:obs}Observations and Data Reduction}

\begin{figure*}
\epsscale{1.0}
\caption{\label{fig:polar} The raw spectrum (ADU), position 
angle (degrees) and linear polarization degree ($\%$) are shown from the 
EFOSC2 spectro--polarimetry observations at day $-$7 (upper panel) and 
day $+$10 (lower panel). Wavelengths are corrected to the host galaxy rest  
frame ($v_{gal}$=2532 \kms, see Sec.~5). The linear polarization and position 
angle data have been binned into 200 channel increments to increase the 
signal--to--noise.} 
\end{figure*}

\begin{figure*}
\caption{\label{fig:spectra}Spectroscopic evolution of   
SN~1998bw from day $-9$ to day $+22$.   
For presentation the spectra have been vertically shifted by arbitrary   
amounts: $-7^d$ ($-0.40$,  $\rm log \; F_\lambda$), $-6^d$ ($-0.60$), $-2^d$   
($-0.95$), $-1^d$ ($-1.10$), $+1^d$ ($-1.30$), $+3^d$ ($-1.45$), $+4^d$   
($-1.60$), $+6^d$ ($-1.70$), $+11^d$ ($-1.75$), $+12^d$ ($-1.90$), $+13^d$  
($-2.10$), $+19^d$ ($-2.30$), $+22^d$ ($-2.45$). Spectra are in the host  
galaxy rest frame ($v_{gal}$=2532 \kms, see Sec.~5).}  
\end{figure*}

\subsection{\label{sec:optspec} Optical Spectroscopy}

Spectroscopic observations were obtained from early to late phases  
using the ESO 1.52m 
(Boller \& Chivens spectrograph), ESO--Danish 1.54m (DFOSC), ESO-3.6m 
(EFOSC2) and the ESO--NTT (EMMI) telescopes. Exposure times ranged 
from 10 minutes near maximum light to several hours at late 
phases. The journal of spectroscopic observations is shown in 
Table~\ref{tab:spec}. 
  
The spectra were reduced using IRAF\footnote{IRAF is distributed by the 
National Optical Astronomy Observatories, which is operated by the Association
of Universities for Research in Astronomy, Inc., under contract to the National
Science Foundation.} packages.  For some of the 
spectra taken at the ESO-Danish telescope, it was not possible to 
remove the fringing in the red since suitable flat fields were not 
available. This gives rise to the high frequency modulation of the 
spectra at wavelengths longer than 7500~\AA.  
  
Particular care was devoted to the extraction of the SN spectra to 
avoid contamination from the host galaxy background. Nevertheless, 
especially in the late phase spectra, the contribution from an 
underlying HII region could not be completely eliminated, and thus 
unresolved narrow lines (H$\alpha$, H$\beta$, [O~II], [O~III]) appear 
in the reduced spectra. 
We emphasize that these features do not show a coherent time evolution,  
but rather depend on seeing conditions and slit position. For this reason we 
conclude that they are not intrinsic to the SN. Finally, there is no 
evidence of continuum contamination from the parent galaxy.  
 
Wavelength calibration was achieved by using arc spectra from He-Ne-Ar 
lamps, while the response curves were obtained via observations of 
spectrophotometric standard stars (Oke 1990; Hamuy et al. 1992). The  
accuracy of the absolute flux calibration was finally checked against the  
broad-band photometry and, when necessary, adjusted accordingly. 
  
On May 6, 1998, two high resolution spectra of SN~1998bw, covering the 
region 3750--7650~\AA, were obtained at the ESO-NTT using EMMI in 
the echelle mode, with a 7$\arcsec$ long and 2\as0 wide slit 
(resolution 1~\AA\/ FWHM at 5900~\AA, i.e. 50 \kms). Order 
definition and extraction, sky subtraction and wavelength rebinning  
were achieved using the IRAF echelle reduction package. The 
orders were merged and the two exposures combined to a weighted mean 
to give the final spectrum which was eventually flux calibrated by 
comparison with a medium resolution spectrum of the SN taken the 
following day (see Table~\ref{tab:spec}). A signal-to-noise ratio of 
about 15 in the region of the NaI D doublet was achieved.

\subsection{IR Photometry and Spectroscopy}  
  
Near IR photometry of SN~1998bw was obtained at three epochs with SofI at 
the ESO NTT (see Table~\ref{tab:irphot}) through the $J$, $H$, and $K$ passbands 
(Bessell \& Brett 1988). In order to allow for a proper sky  
subtraction, the target was observed using the jittering technique. 
The reductions were performed in a standard way within IRAF 
and the fluxes were estimated via plain aperture photometry, since 
at those early epochs the contribution from the host galaxy was negligible.  
The photometric errors were estimated including the nightly zero point  
variations, the aperture correction and the photon shot noise. 
 
Near IR spectra of SN~1998bw were taken at the same epochs and with the same
instrument used for the IR photometry. To cover the entire near IR region, two 
grisms were used: one for the range 0.95--1.64~$\rm\mu m$ and one for the 
range 1.53--2.52~$\rm \mu m$. The observations were made with the 1\as0  
slit. The resulting resolving power, measured from the comparison xenon  
spectrum, was $\lambda/\Delta \lambda\sim$600 for both grisms. 
  
As is standard procedure for IR spectroscopy, the SN was observed at two 
positions along the slit and the telescope was nodded between these two 
positions once every few minutes.  The highly variable night sky is then 
accurately removed by combining all spectra in the appropriate 
way. Atmospheric features have been removed by dividing the extracted 
spectra by the spectra of nearby bright stars that were observed soon 
after the SN: HD~135591 (May 18, Perryman et al. 1997), Hipparcos~106725  
(June 6, Perryman et al. 1997), BS4620 and BS6823 (June 30, McGregory 1995). 
The stellar spectra often contain weak absorption lines 
from hydrogen (Paschen and Brackett series) and helium, which where removed 
before division.  Wavelength calibration was achieved using comparison emission 
line spectra of xenon gas lamps and is accurate to 1--2~\AA. The spectra  
obtained with the two different grisms were then combined into a single  
spectrum.  
 
Relative flux calibration was achieved by multiplying the spectrum by a 
blackbody curve with a temperature appropriate for the star used to remove  
the atmospheric features (HD~135591, T=30,000 K; Hipparcos 106725,  
T=5800 K; BS4620, T=13,000 K, BS6823, T=25,000 K). 
Absolute flux calibration was performed by comparison to the broadband IR  
photometry at the same epochs.

\begin{figure*}
\caption[fig4.eps]{\label{fig:compmax} Comparison between SN~1998bw and   
other SNe at maximum light. Data are from Patat et al. 1995 (1994D),  
Harkness et al. 1987 (1984L), ESO-KP data base (1994I) and Garnavich et al.   
(1997ef, in preparation). All spectra have been corrected to the parent galaxy rest frame adopting the following recession velocities: SN~1998bw, 2532 
\kms\/  (see Sec.~5); SN~1997ef, 3504 \kms; SN~1994I, 463 \kms; SN~1994D, 448 
\kms;  SN~1984L, 1934 \kms\/ (De Vaucouleurs et al. 1991).}  
\end{figure*}

\begin{figure*}
\caption[fig5.eps]{\label{fig:vel} Evolution of the Si~II 6355~\AA\/   
region. The empty circles mark the value which has been assumed to represent   
the photospheric velocity.}  
\end{figure*}

\begin{figure*}
\caption[fig6.eps]{\label{fig:vel2} The Si~II 6355~\AA\/ expansion   
velocity evolution of SN~1998bw compared with SN~1994D (Patat et al. 1995),   
SN~1994I (Millard et al. 1999) and SN~1997ef (Garnavich et al. in  
preparation).  For comparison the photospheric expansion velocities  
deduced from the H$\alpha$ absorption minima of SN~1987A are also   
plotted (Phillips et al. 1988).}   
\end{figure*}

\subsection{Polarimetry}  
  
Spectropolarimetry of SN 1998bw was performed at two epochs (1998 May 4 and 
1998 May 29) using the ESO 3.6m equipped with EFOSC2 in polarimetric mode
(Patat 1999). A Wollaston prism was used with a focal plane mask to isolate 
the ordinary and extra-ordinary ray spectra of object and two sky positions; 
the separation of the two spectra is 20$^{\prime\prime}$. A Half Wave plate 
was used to obtain spectra at  four different position angles of 0\deg0, 
22\deg5, 45\deg0 and 67\deg5. The B300 grism was used, giving a resolution 
of 10~\AA\ over  the wavelength range 3400--7550~\AA. Table~\ref{tab:spec} 
provides a journal of the observations. The airmass was large at  
the time of the observations and the spectral region at $\lambda<$4000~\AA\/ 
is severely affected by differential atmospheric refraction. The HST 
polarized standard star HD~161056 (Turnshek et al. 1990) was observed on 
each night to check the polarization and position angle calibration. 
 
The data were bias subtracted and the ordinary and extra-ordinary ray  
spectra of SN 1998bw were extracted with the local sky subtracted. Statistical  
errors were assigned to each extracted spectrum using the known properties of  
the detector (CCD ESO\#40, Patat 1999). The data were then combined, following
the procedure outlined by Tinbergen \& Rutten (1992) and the total spectrum,  
linear polarization and position angle were computed using dedicated  
programs running in the ESO MIDAS package (see Walsh 1992). The  
polarization was binned into equal wavelength width bins to provide  
polarization errors per bin below 0.1\% over the wavelength range  
4000--7000~\AA. For observations of HD~161056, the measured  
polarization in the V band was 4.3\% at position angle (PA) 76$^\circ$,  
in satisfactory agreement with the standard value of 4.04\% at PA=67$^\circ$, 
given that Turnshek et al. (1990) suggest it may be variable in polarization.

\subsection{Optical Photometry}  
   
Late time optical broad band photometry of SN~1998bw in the Johnson--Cousins  
{\it UBVRI} photometric system (Bessell 1990) was obtained in the phase
range 310--403 days from $B$ maximum. Table~\ref{tab:photo} shows the journal 
of the photometric observations, which were performed with the 
ESO-Dutch 0.9m telescope, equipped with a TEK coated $512 \times 512$ 
pixels CCD (scale 0.44$\arcsec$/pixel), and with the ESO-3.6m telescope, 
equipped with EFOSC2 (Loral/Lesser $2048 \times 2048$ pixels CCD, scale 
0.16$\arcsec$/pixel). The SN is located on a spiral arm and   
superimposed on an HII region, which is clearly visible in Fig.~\ref{fig:map}.
At late phases the background contribution is significant, and the SN 
magnitude cannot be measured using plain aperture photometry. Therefore, after
the standard bias and flat-field corrections, magnitude measurements were 
obtained by means of the point-spread function fitting technique 
implemented in SNOoPY (Patat 1996), a package specifically designed 
for SN photometry within the IRAF environment. 
  
Color terms for the two instruments have been computed using observations  
of standard fields (Landolt 1992). The SN magnitudes have been calibrated   
by means of relative photometry with respect to the local sequence defined by  
Vreeswijk et al. (2001, stars 1 to 10).  
The results of the late photometry here presented are in good agreement  
with those recently published by Sollerman et al. (2000). 
  
\section{\label{sec:spec} Spectroscopic evolution around maximum light}  
  
The spectral behavior of SN~1998bw from day +7 to day +94 has already been 
discussed by Stathakis et al. (2000). The larger wavelength range 
and improved phase coverage of the data we are presenting 
here, allow us to analyze the early phases in more detail and to push the  
discussion further into the late evolutionary stages. For a discussion 
of an independent late phase data set see Sollerman et al. (2000). 
 
The spectroscopic evolution of SN~1998bw around maximum light is 
presented in Fig.~\ref{fig:spectra}. Wavelengths are corrected to the 
parent galaxy rest frame, adopting $v_{gal}=2532$ \kms, as measured 
from the narrow H$\alpha$ emission line (cf. Sec.~\ref{sec:highres}).
Phases are from B maximum, whose epoch (May 10, 1998 UT) was 
estimated from the photometric data presented by Galama et al. (1998b). 
It occurred 14.4 days after the detection of GRB~980425. We adopt these 
conventions throughout the paper.

The general appearance of the spectrum at maximum light is quite unique among 
SNe, even though it is somewhat reminiscent of SN~1997ef (Filippenko 1997c; 
Garnavich et al. in preparation) which has been modeled as a massive SN~Ic  
event (Iwamoto et al. 2000; Mazzali, Iwamoto \& Nomoto 2000). 
This appears clearly in Fig.~\ref{fig:compmax},  
where the spectrum of SN~1998bw is compared to those of other Type I SNe.  
  
Another striking feature is the redward shift with time of most of the 
spectral features, both in absorption and in emission. At early phases
the apparent broad emission--like features do not result from discrete emission
lines, but occur where absorption line optical depths are low and photons
red--shifting in the expanding ejecta, as a result of absorption and scattering
processes, have a higher probability of escaping (see for example Mazzali 2000; 
Pinto \& Eastman 2000). Thus, while the shift of the absorption lines can be 
understood as the effect of the inward recession of the photosphere, that of 
the emission features requires a clearer knowledge of which lines are 
contributing to absorption, and how this changes with velocity (i.e. co-moving 
radius). The effect is also shown in Fig.~\ref{fig:vel}, where the evolution of 
the Si~II~6355~\AA\/ region has been plotted. 
  
In these early phases when the velocity is high, line-blending is 
particularly severe. The modeling presented by Iwamoto et al. (1998) 
suggests that the main features are due to lines of Si~II, O~I, Ca~II 
and Fe~II.  The time evolution of the expansion velocity at the 
photosphere, as deduced from the minimum of the absorption trough of 
the Si~II~6355~\AA\/ line, is shown in Fig.~\ref{fig:vel2}.  For 
comparison, the corresponding values measured for the Type Ia SN~1994D, 
the Type Ic SN~1994I\footnote{The data for SN~1994I are from the
ESO Key Project on Supernovae Data Base (Turatto et al. 1990), hereafter ESO--KP.} 
and the peculiar Type Ic SN~1997ef are also 
plotted. SN~1998bw shows a sudden break in the expansion velocity 
decline rate around day 15 and another at day 27. While changes in the 
slope are to be expected as a result of changes in optical depth at 
line center, the reasons for sudden changes are not clear. There is no doubt
that the velocity decline rate changes significantly; what is less clear, in
view of measurement uncertainties, is how suddenly they occurred.
  
The velocity deduced from the Si~II~6355~\AA\/ line is about $30,000$ \kms\/ 
at day $-$7, and decreases to about 8000 \kms\/ at day $+$22. These values are 
exceptionally high, for any SN, as can be seen in Fig.~\ref{fig:vel2}.  
SN~1994I shows a trend which is very similar to that of SN~1998bw, although  
the velocities are systematically lower, while SN~1997ef deviates  
significantly after day 25 (Fig.~\ref{fig:vel2}). Phases for SN~1997ef are  
uncertain because maximum light was not recorded.

\begin{figure*}
\caption[fig7.eps]{\label{fig:irspec} The IR spectroscopic evolution of   
SN~1998bw from $+8$ to day $+51$.  
For presentation the spectra have been vertically shifted by an arbitrary   
amount: $+33^d$ ($-0.80$, $\rm log \; F_\lambda$), $+52^d$ ($-1.70$).  
The vertical dotted line is placed at the wavelength  
of the absorption trough's minimum of He~I 1.083~$\rm \mu m$.  
The increased noise level in the wavelength ranges 1.35--1.45~$\rm \mu m$   
and 1.8--1.9~$\rm \mu m$ is due to high atmospheric absorption. 
Spectra are in the host galaxy rest frame.}  
\end{figure*}  

\begin{figure*}
\caption[fig8.eps]{\label{fig:hevel}Upper panel: IR spectroscopic  
evolution of SN 1998bw from day $+8$ to day $+51$ in the region of He~I  
1.083~$\rm \mu m$. The vertical dashed line marks the expansion velocity  
measured for He~I 1.083 $\rm \mu m$ on day $+8$. 
Lower panel: region of He~I 2.058~$\rm \mu m$ at day +51. The solid line 
is a smoothed version of the spectrum. A  350 \AA\/ wide boxcar filter was 
used.}  
\end{figure*}  

\begin{figure*}
\caption[fig9.eps]{\label{fig:he} Optical and IR spectra of SN~1998bw   
at comparable phases.   
Line identifications from spectral modeling are plotted for the most   
prominent emission features (top) and for the He~I lines (bottom). The He  
marks are placed at the expected absorption positions for an expansion 
velocity of 18,300 \kms.}  
\end{figure*}

\section{\label{sec:polar} Polarimetry}  
  
In order to investigate any possible asphericity in the explosion of SN  
1998bw, spectro--polarimetry measurements were performed at two epochs  
(day $-$7 and day +10, see Tab.~\ref{tab:spec}). The polarization averaged over  
the range 4000--7000~\AA\ was 0.7\% and 0.5\% at the two epochs respectively.  
Figure \ref{fig:polar} shows the total spectrum (raw, not flux calibrated),  
the polarization binned into 200 channel 
bins and the position angle of the polarization vector projected on the sky 
at both epochs. The formal errors per bin are about 0.04\% but the 
error on the polarization binned over the optical range is 0.1\% 
given that the measurements were not of high photometric  
quality and the discrepancy on the measurement of the polarized standard. 
It appears that the polarization spectrum at day $-$7 may have a significant  
increase to the red which is not in evidence at the later epoch.  
In terms of the integrated polarization over the  4000--7000~\AA\ range,  
there may be a marginal change in the polarization between the two epochs,  
but no great significance is attached to this result. 
 
The interstellar polarization 
in the direction of the host galaxy can be estimated by the measurement of 
stellar polarization along a nearby sight line. Kay et al. (1998) 
measured the linear polarization of HD~184100 at 0.75\% in PA 176.5. 
Correcting the linear polarization results by this value of interstellar 
polarization results in 0.6\% at PA 80$^\circ$ and 0.4\% at 67$^\circ$ at 
day $-$7 and +10 respectively. The polarization value is not very different  
from the value of 0.53\% measured by Kay et al. (1998) at day $\sim$42 and 
is most probably intrinsic to the SN, although a dusty medium in the 
SN parent galaxy cannot be ruled out. However the position angles differ  
from the Kay et al. value of 49$^\circ$, although the errors on these  
earlier determinations are larger (at least 5$^\circ$). Despite its  
association with a $\gamma$--ray burst, these polarization values are similar  
to those reported for other core-collapse SNe (Wang et al. 1996). 
  
The small degree of polarization at optical wavelengths can 
be explained in terms of a moderate departure from sphericity (axial 
ratio $<$2:1, H\"oflich 1995, H\"oflich et al. 1999) either in the 
photosphere or in the outer scattering envelope when the line-of- 
sight is not coincident with an axis of symmetry. Unfortunately, it is 
not possible from polarization measurements alone to decode both the 
shape of the object and the viewing angle and therefore the small polarization 
values reported here could indicate large departures from sphericity 
but viewed close to an axis of symmetry, or small departures viewed 
well away from the symmetry axis. Net polarization could also result 
from a spherical envelope in which, as a result of large scale 
clumping, one hemisphere had a surface brightness different from that 
of the other hemisphere.  
Delayed light echoes could also be responsible for the observed 
polarization, provided that the time delay is comparable to the time scale 
of the SN luminosity variation. 
  
We note that no polarization, neither circular nor linear, was 
detected in the radio and this has been interpreted as the signature 
of a spherically symmetric blast wave (Kulkarni et al. 1998). This 
apparent discrepancy might suggest that the radio and the optical 
radiation were generated in regions of different geometry.

\begin{figure*}
\caption[fig10.eps]{\label{fig:spectra2} Spectroscopic evolution of   
SN~1998bw from day $+29$ to day $+376$.   
The phases have been computed from B maximum light (May 10, 1998).   
For presentation the spectra have been vertically shifted by arbitrary   
amounts: $+45^d$ ($-0.20$), $+52^d$ ($-0.55$),  
$+64^d$ ($-0.95$), $+73^d$ ($-1.20$),$+94^d$ ($-1.50$), $+125^d$ ($-1.70$),   
$+201^d$ ($-1.90$), $+337^d$ ($-2.00$), $+376^d$ ($-2.70$). Narrow emission  
lines visible in the late time spectra originate in the parent galaxy. 
Spectra are in the host galaxy rest frame.}  
\end{figure*}  

\begin{figure*}
\figcaption[fig11.eps]{\label{fig:complate} Comparison between SNe 1992A   
(Ia, ESO-KP data base), 1998bw, 1996N (Ib, Sollerman et al. 1998) and  
1996aq (Ic, ESO-KP data base, unpublished) at late phases.  The vertical  
dashed line is placed at the rest-frame wavelength of Mg~I]~4571~\AA.}  
\end{figure*}

\section{\label{sec:highres} High Resolution Spectroscopy}  
  
To search for possible narrow emission lines produced in the 
circumstellar environment of the progenitor and for interstellar 
absorption features, we obtained high resolution spectra of SN~1998bw 
around maximum brightness.  With the exception of a narrow H$\alpha$ 
emission line centered at 2532 \kms\/ and probably arising from the underlying  
HII region (see also Sec.~\ref{sec:optspec} and 
Tinney et al. 1998), neither the usual NaI~D absorption lines, which would  
signal intervening interstellar dust, nor narrow emission lines 
were detected. However, the signal-to-noise ratio we achieved 
($SNR\sim20$ at H$\alpha$ and $\sim15$ at NaI~D) is insufficient to 
exclude the presence of lines, both in emission and absorption, with 
EW$\leq 0.2$~\AA. Using the relation recently calibrated by Benetti et 
al. (in preparation) and assuming that the total EW of the NaI~D doublet is  
EW$ \leq$ 0.4~\AA, one can estimate an upper limit for the  
extinction, $A_V\leq$0.2. There are two values of $A_V$ 
available in the literature, i.e. $A_V$=0.05 (Burstein \& Heiles 1982) and
$A_V$=0.2 (Schlegel, Finkbeiner \& Davis 1998). Both estimates are compatible 
with our upper limit, but we will assume $A_V$=0.2 throughout this paper 
due to its more recent determination from COBE/DIRBE and IRAS/ISSA maps.

\section{\label{sec:helium} IR Spectra: Helium and Other Elements}  
  
The IR spectra are plotted in Fig.~\ref{fig:irspec}. The most 
prominent feature is the broad emission at 1.08~$\rm \mu m$, which has an 
associated P-Cygni structure with a hint of more than one 
component in absorption. The main component of the emission and the 
absorption minimum at day +8 strongly suggest that this feature is due 
to He~I 1.083~$\rm \mu m$ with a velocity of 18,300$\pm$700 \kms. There is
nevertheless the possibility that transitions from other ions are affecting
this profile.
The absorption does not seem to shift red-wards with time as shown in the upper 
panel of Fig.~\ref{fig:hevel}. This is to be expected if the helium were  
confined to an outer higher velocity layer either optically thick or thin
but restricted to a narrow velocity range. A more extended layer which had
become optically thin on or before the date of the first IR spectrum
could also in principle give rise to this constancy of velocity. The first
interpretation is reinforced by the fact that at the 
same phase (i.e. 22.4 days after the explosion),  
Si~II 6355~\AA\ had a much smaller velocity ($12,000$ \kms)  
which may more closely indicate the photospheric velocity, 
at least up to day $+22$ (see Fig.~\ref{fig:vel2}). Since the color temperature
of the envelope at those phases is not sufficiently high to radiatively or
collisionally excite the required levels of He~I, one may conclude that 
excitation is caused by non--thermal electrons produced by $\gamma$-rays 
coming from the radioactive decay of $^{56}$Co. This 
was predicted by Chugai (1987) for SN~1987A, and shown to be effective  
by Graham (1988), Lucy (1991) and Mazzali \& Lucy (1998) in various SNe,
after Harkness et al. (1987) had identified He~I lines by postulating large
departures from LTE to explain the abnormal strengths.
 
He~I has another transition (singlet series) in the near IR, namely at 
2.058~$\rm \mu m$ ($2s^{1}S-2p^{1}P^{0}$). Careful inspection shows that in  
all three spectra a weak broad P--Cygni feature is present at this wavelength  
(see Fig.~\ref{fig:irspec}). While this confirms the presence of He~I in the
spectra, because this spectral region is affected  
by strong telluric absorption, any line strength measurements should be
treated with some caution.
In the lower panel of Fig.~\ref{fig:hevel} the region centered on  
2.058~$\rm \mu m$ is plotted for day +51, i.e. when this feature appears to be 
more pronounced. Even though the signal--to--noise is 
poor, a P--Cygni profile is visible and from the minimum of the absorption 
through an expansion velocity of 13,000 $\pm$ 2000 \kms\/ is deduced, while the 
measured intensity ratio He~I 1.083/He~I 2.058 is about 30. 
 
Since these He~I lines show P--Cygni profiles associated with strong
continuum radiation, it is instructive to follow the qualitative results
obtained by an elementary SN model in the Sobolev approximation 
(see for example Jeffery \& Branch 1990). In that framework, the absolute emission 
line intensity depends not only on the line optical depth, but it is also 
proportional to the photospheric continuum intensity at the line wavelength.
Thus it is not surprising that the emission component of He~I 
1.083~$\rm \mu m$ is much greater than that of He~I 2.058~$\rm \mu m$. 
There is in fact roughly a factor 10 difference in the continuum flux at the two
wavelengths, and therefore the measured intensity ratio translates into an 
equivalent width ratio $EW(He~I 1.083)/EW(He~I 2.058)\sim$3.
The theoretical work by Lucy (1991) on the non--thermal excitation of helium in 
Type Ib SNe predicts a ratio close to one for these two He~I features, whereas 
the ratio between the A--values is 5.2 (Martin 1987).

An additional complicating aspect to consider is that the He~I 1.083~$\rm \mu m$ 
line originates from the metastable triplet ground state which can give rise to 
resonance scattering not expected for the He~I 2.058~$\rm \mu m$ line, whose lower
level is metastable but whose upper level can be depopulated by an allowed 
transition to the singlet ground state. Moreover, the intensities of the two lines
depend also on the fraction $\epsilon$ of collisional to total depopulations of the 
upper level of the transition. Hence, if one takes also into account that the 
optical depths of the two lines can be different, it is not surprising that 
the observed line ratio deviates from the expectations based on Einstein 
coefficients alone.

We note that the IR spectra of SN~1987A have shown a strong He~I 
1.083~$\rm \mu m$ line starting at day 110 (Meikle et al. 1989), from which an
expansion velocity of 5000~\kms\/ was deduced. The He~I 2.058~$\rm \mu m$  
line was also detected with an expansion velocity of 1500~\kms. Meikle et al.
(1989) suggested that this is probably due to the fact that He~I 2.058~$\rm \mu m$
was optically thin. For SN~1987A the ratio He~I 1.083/He~I 2.058 
in the phase range 110--349 days ranged from 23 to 50. In  particular, on day 112 
it was about 35; once one has taken into account the different continuum level 
at the two wavelengths, this turns into an equivalent width ratio of about 6. 

Very few other SNe have been observed in the 2~$\rm \mu m$ region.
The Type II SN~1995ad and Type Ic SN~1997B have both shown a strong He~I 
1.083~$\rm \mu m$ feature without any evidence of the He~I 2.058~$\rm \mu m$ line 
(Clocchiatti et al. 2001). He~I 1.083~$\rm \mu m$ was identified also in the near
IR spectra of the Type IIn SN~1998S (Gerardy et al. 2000) and in that case,
even though very faint, the He~I 2.058~$\rm \mu m$ line was detected at 95 days 
after $V$ maximum.

One possible difficulty with the identification of the 1.083~$\rm \mu m$ feature 
as He~I comes from the apparent absence of the He~I lines in the 
optical spectra at comparable phases. In Fig.~\ref{fig:he} we plot the 
optical spectra at day +6, +29 and +52, i.e. as close in time as possible to 
the IR spectra.  At all three epochs, the spectrum is predominantly an 
absorption spectrum.  Some of the strongest features tentatively 
identified using spectral synthesis are marked (top). We also marked 
the expected position of the He~I absorption lines if they all had a blue-shift 
of 18,300 \kms\/ (bottom).  The strongest optical line 
of He~I should be 5876~\AA, but this is blended with Na~I~D. He~I 5876 
\AA\ is clearly not present as a distinct individual feature at 18300 
\kms. There is no unambiguous sign of any other possible He~I optical lines.  
This does not necessarily contradict the identification of the 
1.083~$\rm \mu m$ feature as He~I, because the optical lines can be 
suppressed relative to the IR ones if non-thermal excitation is at work 
(Mazzali \& Lucy 1998). 
The large expansion velocity of SN~1998bw might also dictate greater 
blending effects in the optical region where the number of lines per unit 
wavelength is much larger than in the IR.  
  
The optical He~I lines are at least weaker than in the Type Ic SN~1994I. This 
SN also showed the He~I 1.083~$\rm \mu m$ line (Filippenko et al. 1995) and 
possibly traces of He~I contamination in the optical (Clocchiatti et al. 1996).   
Millard et al. (1999) have modeled SN~1994I's spectra and could fit  
the feature near 1.08$~\mu m$ using a blend of CI 1.0695~$\rm \mu m$ and He~I 
1.083~$\rm \mu m$, the latter being detached at 18,000 \kms. 
This velocity is similar to what is seen in SN~1998bw.  However, they were
able to obtain an alternative fit using only Si~I lines.   
Another possibility is that the feature at 1.08$~\mu m$ is due to 
Mg~II 1.091$~\mu m$, as suggested by Mazzali \& Lucy (1998) for the Type Ia 
SN~1994D.  

Determining the amount of He in the ejecta is very important for 
constraining the nature of the progenitor and its evolutionary stage 
at the time of the explosion. Detailed spectral modeling, including 
non--thermal effects, is necessary for this purpose. Inclusion of 
helium in the exploding progenitor object may even affect our 
quantitative conclusions concerning the ejected mass, mass of radioactive
material, kinetic energy and mixing of the ejecta. 
  
As the SN ages the 1.083~$\rm \mu m$ emission feature develops a second, 
redder component (cf. Fig.~\ref{fig:irspec}). Possible identifications  
are O~I 1.129~$\rm \mu m$ and 
Na~I 1.138, 1.140~$\rm \mu m$. The latter seems to be possible since its 
lower level is the upper level of the NaI D line transition which 
would be well populated by resonance scattering. The Na~I 
identification implies a velocity close to that of Si~II 6355 \AA\/ at the 
same phase and consistent with the Na~I D line absorption. The 
presence of O~I 7771~\AA\/ in the optical is only a weak argument in favor 
of the O~I 1.129~$\rm \mu m$ identification, since this latter line 
originates from a higher level. 
  
\begin{figure*}
\figcaption[fig12.eps]{\label{fig:oxprof} Upper and central panel: 
[O~I]~6300,6364~\AA\/ line profile at two different epochs for SN~1998bw 
(thick line) and SN~1996N (thin line, Sollerman et al. 1998). The vertical 
dotted line is placed at the position of the unidentified feature 
(see text). Velocities are computed from the 6300~\AA\/ doublet component.
Lower panel: Comparison between the profiles of [O~I]~6300,6364~\AA\/ (thick
line) and MG~I]~4571~\AA\/ (thin line) for SN~1998bw at +376 days.}  
\end{figure*}  

\begin{figure*}
\figcaption[fig13.eps]{\label{fig:linevol} Evolution of Mg~I]~4571~\AA\/   
(right panel) and [O~I]~6300,6364~\AA\/ (left panel) line profiles.}  
\end{figure*}  

\begin{figure*}
\figcaption[fig14.eps]{\label{fig:fluxvel} Upper panel: FWHM expansion   
velocity deduced for Mg~I]~4571~\AA\/ and [O~I]~6300,6364~\AA. Lower panel:   
evolution of line luminosities. Dotted and dashed lines represent linear  
fittings to the data; line lengths indicate the phase range which has 
been used for the linear regression. The box in the upper right corner of the 
lower panel shows the region where the flux for [Fe~II] 5215~\AA\/ has been 
estimated and the position of the two continuum levels (dashed and dotted  
lines). See text for more details.}  
\end{figure*}

\section{\label{sec:nebular} Entering the Nebular Phase}  
  
The spectral evolution of SN~1998bw towards and during the nebular 
phase is presented in Fig.~\ref{fig:spectra2}, which shows all our 
spectra taken between day +29 and day +376. The transition from an 
absorption to an emission spectrum is slow and subtle and is marked 
by a decrease in the continuum. 
   
While the evolution of SN~1998bw in the range 5500--9000~\AA\/ is similar  
to that of SN 
Ic events such as SN~1987M (Filippenko et al. 1990), the expansion 
velocities are larger and the region between 4000 and 5500~\AA\/ is 
dominated (at least until about day +200) by a wide bump to which Fe~II 
transitions probably contribute significantly. The spectroscopic 
differences appear clearly in Fig.~\ref{fig:complate}, where SN~1998bw 
is compared to the Type Ia SN~1992A, the Type Ib SN~1996N and  
the Type Ic SN~1996aq. From Fig.~\ref{fig:complate} the late nebular spectrum of
SN~1998bw appears to mostly resemble a Type Ic SN, in this case 
SN~1996aq. Mg~I]~4571~\AA\/ and [O~I]~6300~\AA\/ are present in both cases  
but not in the Type Ia SN~1992A. SNe Ia in general show strong [Fe~II] and 
[Fe~III] emission features not apparent in the same form in the two Type Ic 
SNe. In fact, the strongest spectral resemblance of SN~1998bw at 
nebular phases is to the peculiar SN~1985F and the Type Ic SN~1987M  
(Filippenko et al. 1990). Qualitatively the appearance of Mg~I and O~I is 
consistent with what is expected from the nucleosynthesis model of Iwamoto 
et al. (1998), and their absence in models of Type Ia proposed by Nomoto et 
al. (1997).
  
All these facts support the general idea that SN~1998bw is related to 
SNe~Ib/c. It might be regarded as an extreme case among these objects, 
having large kinetic energy, ejecta mass and ejected mass of synthesized  
$^{56}$Ni, while SN~1997ef could represent a less extreme case closer in
properties to the known SNe Ic (see also Iwamoto et al. 2000; Mazzali et al. 
2000; Stathakis et al. 2000, Branch 2001).

At late epochs the most prominent spectral feature
is the [O~I]~6300,6364~\AA\/ doublet. Fig.~\ref{fig:oxprof} (upper and central panel) 
reveals that this line has a complex profile, showing an apparently blended
emission component blue shifted by approximately 2300 \kms\/ with respect to 
the 6300~\AA\/ doublet component. It is present at 201 and 376 
days, and a similar feature appears to be present in the spectra of 
SN~1996N at comparable phases. This therefore points to a common 
origin in the form of an as yet unidentified line (centered at $\sim$6250~\AA), 
or a similar structure (a clump?) in the line emitting region. 

The [O~I]~6300~\AA\/ line profiles are sharply peaked in the nebular phase, as 
are those of Mg~I]~4571~\AA\/ albeit a little less clearly. Uniform emission 
throughout the envelope should produce a parabolic profile, and 
therefore we can conclude that stronger emission is occurring at lower  
velocities in regions closer to the center (see also Maeda et al. 2000).
A similar effect is apparent in other SNe such as the Type Ic SN~1987M and 
peculiar SN~1985F (Filippenko et al. 1990) where it is also evident in emission 
lines of other ionic species.  
  
Since Mg~I]~4571~\AA\/ seems to be the only single line in the 
spectrum with a sufficient signal-to-noise ratio, it is of some 
interest to study its behavior with time (see Fig.~\ref{fig:linevol}, 
right panel). At early stages prior to 100 days the apparent emission 
feature must be dominated by some as yet unidentified feature because 
the peak is displaced bluewards by 3500 \kms\/ and gradually reaches 
the rest wavelength of Mg~I]~4571~\AA\/ at day 125. From this phase onwards 
there is an excess in the blue wing causing an asymmetry, which 
becomes more marked as time passes by (see also Fig.~\ref{fig:spectra2}). 
One explanation for this blue excess after day 201 is that it is the remnant 
of the unknown feature that dominated at early phases. It cannot be caused 
by the presence of Mg~I]~4562~\AA\/ which only reaches strengths comparable 
to that of Mg~I]~4571~\AA\/ at densities lower than a few 10$^{4}$  
cm$^{-3}$ and in any case the separation is too small. A more interesting and
plausible explanation for this blue component of Mg~I]~4571~\AA\/ is that it is
the counterpart of the blue component of [O~I]~6300~\AA\/ mentioned above. The
velocities and the profiles are similar (Fig.~\ref{fig:oxprof}, lower panel). 
That being so, it suggests that we are seeing two components of the oxygen and 
magnesium mass distribution in the envelope.

We estimated the expansion velocity of the Mg nebula fitting the  
Mg~I]~4571~\AA\/ line profile. The results are plotted in the upper panel of  
Fig.~\ref{fig:fluxvel}. We emphasize that, especially at very late phases,  
the fitting is not satisfactory and the values 
plotted are only indicative. A comparison with other SNe of similar 
Type is interesting. The FWHM of Mg~I]~4571~\AA\/ in the latest spectra 
of the Type Ib SN~1983N (225 days, Gaskell et al. 1986) and the Type Ic  
SN~1996aq (270 days) was about 5000$\pm$300 \kms\/ and 7000$\pm$800 \kms\/ 
respectively. A similar value (6100 \kms) was reported by Sollerman et al. (1998) 
for the Type Ib SN~1996N at about 220 days. These values are to be compared with 
9800$\pm$500 \kms\/ measured for the same line in SN~1998bw at 201 days past 
maximum. However, this value reduces to 7700$\pm$800 \kms\/ if we determine the 
velocity using only the red half of the profile. Thus it seems that when  
account is taken of the somewhat different phases at which measurements  
were made for these SNe, not very significant differences are noticeable. 
  
A similar analysis is more complicated for the [O~I] doublet, 
both because of its complex profile (cf. Fig.~\ref{fig:oxprof}) and 
a strong contamination by the Si~II 6355~\AA\/ line on the red wing at 
phases earlier than 3 months (see Fig.~\ref{fig:linevol}, left 
panel). We tried to estimate the expansion velocity of the O~I 
nebula using the blue wing of the observed profile and hence computing 
the FWHM of the 6300~\AA\/ component alone. The results are plotted in 
the upper panel of Fig.~\ref{fig:fluxvel}, which shows that the 
velocities are similar to those computed from Mg~I].
  
Finally, we have measured the flux of the Mg~I] and [O~I] lines to 
look for possible deviations from exponential decay. The results are 
plotted in the lower panel of Fig.~\ref{fig:fluxvel}, where it appears 
clearly that at late phases the flux in these lines decreases exponentially.  
The measured decline rates are 0.0185 and 0.0170 mag day$^{-1}$ for the two 
features, respectively. These values are quite similar to those 
measured for the broad band photometry (cf. Sec.~\ref{sec:lcurves}). 
It should be noted here that Si~II~6355~\AA\/ line emission   
probably dominates the [O~I] 6300,6364~\AA\/ lines up to day 94 although there is the 
possibility that if [O~I] is optically thick and the expansion 
velocities significantly exceed the doublet separation, most of the 
[O~I] photons would be redshifted and confused with those belonging to 
the Si~II~6355~\AA\/ transition.  
 
The unidentified blue--shifted component of 
Mg~I] will clearly dominate the photometry of this feature prior to 
day 125 and still contribute afterwards. The fact that Mg~I] 
appears to reach the exponential decay line before [O~I]  
(cf. Fig.~\ref{fig:fluxvel}, lower panel) could be due 
to the nature of these blending effects. Also other considerations may 
be important such as the fact that the critical density for Mg~I] is 3 
orders of magnitude higher than that for [O~I] and the expanding 
envelope results in the former critical density being reached well 
before the latter. Also, since Mg~I has a much lower ionization energy 
than O~I, emission of these different lines may be coming from 
different regions. Finally models of massive stars (Woosley, Langer \&  
Weaver 1995) show that magnesium occupies a more confined 
radial distribution than oxygen. This is the case not only for extremely 
massive stars but also for 13--15 $M_{\odot}$ stars (cf. Fig.~1 of Nomoto 
et al. 1995). In both circumstances, the region where O exists without Mg  
is mostly He--rich, the He mass fraction being $\sim$0.1. 
  
Unfortunately, it is not possible to make a similar analysis for the 
presumed Fe lines and particularly the feature near 5215~\AA. In fact 
even the identification of this feature is in doubt. One promising 
possibility is that it is due to the [Fe~II] multiplet 19 with 
additional components due to the [Fe~II] multiplets 16,17,18. [Fe~II] multiplet 6
may then contribute to the blend near Mg~I]~4571~\AA. Nevertheless, 
there is no unequivocal evidence for the presence of other multiplets of 
[Fe~II]. Another possibility suggested by Filippenko et al. (1990) for SN~1985F and 
SN~1987M is a combination of Fe~II multiplets 35,42,49. 
Contributions from [Fe~III] multiplet 1 seems a more remote 
possibility because some important lines are not present. What we can 
state here is that the most prominent feature, 
the one at 5215~\AA, declines faster than those of Mg~I] and [O~I].  
The flux measurement is strongly hampered by the uncertain position of the 
continuum and the possible presence of another feature in the red wing 
(see Fig.~\ref{fig:spectra2}). To reduce the effect of these problems we  
computed the fluxes by integrating the spectra in the range 5080--5680~\AA\/  
and crudely assuming the continuum level. To estimate the  
uncertainty involved we chose the continuum at two extreme positions. One 
is the average value in the range 5680--5700~\AA, where the flux  
drops to a minimum value, and the other is the level used in computing the  
flux of the Mg~I]~4571~\AA\/ measurements. 
 
The results are shown in the lower panel of Fig.~\ref{fig:fluxvel}, where the 
upper and lower extremes of the error bars represent the values obtained 
using the low and high continuum respectively. 
The conclusion is that the 5215~\AA\/ feature fades at a rate of 
0.021$\pm$0.001 mag day$^{-1}$, i.e. 0.0025$\pm$0.001 and  
0.004$\pm$0.001 mag day$^{-1}$ faster than [O~I] and Mg~I] respectively.  
The faster decline of 5215~\AA\/ is also clearly visible, if one compares its 
intensity with the one of the adjacent Mg~I]~4571~\AA\/ after normalization 
to its peak (see Fig.~\ref{fig:mgnorm}).  
This fact might also suggest that the 5215~\AA\/ feature does  
not trace the bulk of the Fe in the envelope since one would expect the  
abundance of Fe to be increasing with time as a result of the radioactive  
decay of $^{56}$Co. This might be partially compensated by the decreasing temperature
and density of the ejecta.
 
\begin{figure*}
\figcaption[fig15.eps]{\label{fig:mgnorm} The region of Mg~I] and [Fe~II] at 
125, 201 and 376 days (from top to bottom). The spectra have been normalized 
to the Mg~I]~4571~\AA\/ line peak. For presentation, the narrow emission 
lines emitted by the underlying HII region in the range 4800--5000~\AA\/  
have been cut out and the spectra have been smoothed using a 20\AA\/ wide 
boxcar filter .}  
\end{figure*}

Another point concerns the expansion velocity deduced from this 
feature. Again the measured values are quite uncertain, but at day 201 
the FWHM velocity is about 10,900 
\kms\/ and hence higher than those estimated for Mg~I]~4571~\AA\/ (9800 \kms) 
and [O~I]~6300,6364~\AA\/ (7600 \kms).  However, if the 5215~\AA\/ 
feature results from a blend of multiplet lines of whichever type, 
this would most certainly lead to an overestimate of the expansion 
velocity since the separation of individual lines of the most likely 
multiplets is of the order of thousands of \kms. Here we only note 
that recent Chandra X--ray observations of Cassiopeia A, which is 
consistent with the remnant of a massive star explosion 
(e.g. Fesen et al. 1987), show that significant amounts of Fe are 
present at high velocities, comparable to those of lighter elements 
such as oxygen (Hughes et al. 2000). This implies that a relevant 
fraction of the material synthesized in the core has mixed into the 
outer regions during the SN explosion. 
  
Finally, we have compared SN~1998bw with the Type IIb SN~1993J
and the Type Ib SN~1996N at about one year after maximum light 
(Fig.~\ref{fig:sn96n}). 
The resemblance of 
the three objects is evident. In particular, SN~1993J and SN~1998bw 
are strikingly similar, the only deviation being the broad H$\alpha$, 
which is missing in SN~1998bw; apart from this, line ratios and 
expansion velocities are definitely comparable. SN~1996N also has very 
similar line widths, while the luminosity ratio between [O~I] and all 
other features is smaller and the blue [Fe~II] bump is less pronounced. 
  
Despite the great peculiarity shown in the early phases, at one year  after  
maximum SN~1998bw is spectroscopically indistinguishable from known Type Ib's.  
Even the high expansion velocities measured during the first six months have  
slowed down to the values which are typical for other Type Ib SNe  
($\sim$5000 \kms). But the much higher ejected mass estimated by the models  
and the high luminosity, which persists also at these advanced phases 
(cf. next section), confirm that this event was hyper--energetic.  
  
\begin{figure*}
\figcaption[fig16.eps]{\label{fig:sn96n} Comparison between Type IIb SN  
1993J  (Barbon et al. 1995, dotted line), Type Ib SN~1996N (Sollerman et al.  
1998) and SN~1998bw at about one year after maximum light. Spectra have been   
normalized to the [O~I]~6300,6364~\AA\/ peak and arbitrarily shifted for   
presentation.}  
\end{figure*}

\section{\label{sec:lcurves} Broad Band and Bolometric Light Curves}  
  
The photometric data presented in this paper and those published by Galama et al. 
(1998b), McKenzie \& Schaefer (1999) and Sollerman et al. (2000) can be collected in 
a unique data set to study the late phase behavior of the light curves, as shown in 
Fig.~\ref{fig:lclate}. Fig.~\ref{fig:lclate} shows that the SN luminosity follows
an exponential luminosity decline up to about 300 days after 
maximum light, while a clear flattening is present for the latest observed  
epochs, especially in the $V$ band.  This is clearly shown in 
Tab.~\ref{tab:slopes}, where we have reported the decay rates computed via
linear least squares fit to the data in the two phase ranges 40--330 and
300--490 days from B maximum. The values for the earliest data are consistent,
within the estimated errors, with those computed by McKenzie \& Schaefer (1999) using 
data in the range 47--171 days ($\gamma_B=0.0141\pm0.0002$, 
$\gamma_V=0.0184\pm0.0002$, $\gamma_I=0.0181\pm0.0002\; \rm mag \; \rm day^{-1}$). 

In any case, these slopes deviate clearly from the values expected 
from the $^{56}\rm Co \rightarrow ^{56}$Fe radioactive decay in the 
case of complete $\gamma$-ray trapping (0.0098 mag day$^{-1}$), as already 
noticed by McKenzie \& Schaefer (1999). On the other hand, the 
measured slopes are similar to those reported by Sollerman et al.
(1998) for the Type Ib SN~1996N in the phase range 180--340 days 
($\gamma_V=0.0167\pm0.0023$, $\gamma_R=0.0172\pm0.0010$, 
$\gamma_I=0.0193\pm0.0024\;\rm mag \; \rm day^{-1}$). These values are even  
higher than those typical for the late phase data of Type Ia SNe. If 
$^{56}\rm$Co decay is powering the late light curve of SN~1998bw, 
there must be a fairly strong leakage in the $\gamma$--ray deposition, 
as for SN~1996N (Sollerman et al. 1998).  Interestingly,  
fast and slow decliners seem to exist within the Type Ic class, as 
pointed out by Clocchiatti \& Wheeler (1997). The range spanned by 
late phase decline rates of SNe Ib/c is quite broad: 
$\gamma_B$ ranges from 0.006 (SN~1990B, Piemonte 2001) to 0.022 mag  
day$^{-1}$ (SN~1990I, ESO--KP data base). 
At later phases ($t\geq$300 days) a clear flattening of the light curve is visible
in all passbands, even though the conclusion about $B$ band is somewhat hampered
by the lack of data at phases later than 403 days. The decline rate 
variation is 0.0020$\pm$0.0013, 0.0079$\pm$0.0010,
0.0036$\pm$0.0020 and 0.0043$\pm$0.0013 mag day$^{-1}$ in $B$, $V$, $R$ and $I$
respectively.

In Fig.~\ref{fig:lcmodel}, we 
present the UVOIR bolometric light curve of SN~1998bw, which was constructed using 
all available data by integrating the flux in the optical and near IR photometric 
bands.  
Magnitudes were converted to fluxes using the calibrations of Bessell (1979)  
and Wilson et al. (1972). The measured fluxes were then transformed to  
luminosities by adopting a distance modulus $\mu=32.89$  
($d=37.8$ Mpc; $H_0=65 \; \rm km \; \rm s^{-1} \; \rm Mpc^{-1}$),  
and an extinction A$_V = 0.2$. We emphasize that {\it JHK} photometry is available 
only for the early phases and that its contribution to the total flux is quite  
significant, ranging from $\sim 25\%$ on day 22.4 (from GRB) to $\sim 35\%$  
on day 65.4. For later epochs we simply assume that the fractional IR contribution  
remained constant at $35\%$ as for the last measurement. Clearly this is a 
major source of uncertainty for the late UVOIR bolometric light curve. 
 
After the maximum peak, where the SN reaches a luminosity of $\sim10^{43}$ 
erg s$^{-1}$, the light curve settles on the exponential decay tail starting  
with day +40, with a decline rate $\gamma$=0.0176$\pm$0.0002 mag day$^{-1}$.  
Then, after day +300, the bolometric light curve appears to flatten, gradually 
deviating from the slope defined by the data in the range 40--200 days (dotted line)  
so that at +490 days the difference is about 1.4 mag, which is well  
outside the error bars (cf. Fig.~\ref{fig:lcmodel}).  
These uncertainties were estimated taking into account the presence of possible 
contamination from the unresolved  underlying HII region, whose contribution is 
difficult to subtract (see below).  

The comparison of the bolometric light curve of SN~1998bw with the model of  
Iwamoto et al. (1998) is also presented in Fig.~\ref{fig:lcmodel}. Although  
the early phases (t$\leq$ 60 days) are well reproduced, the observed tail  
deviates from the theoretical prediction. The model must be highly energetic  
to accommodate the early light curve and spectra. Therefore at later phases  
$\gamma$--rays should escape quite efficiently resulting in a rapid decline. 
Eventually most $\gamma$-rays escape from the ejecta, and only positrons from  
the $^{56}$Co decay contribute to the light curve (the masses of $^{57}$Co and  
$^{44}$Ti in the model are small). 
Indeed, after day $\sim$200 the calculated decline becomes slower, and it  
approaches the decay rate of $^{56}$Co around day 400 (Iwamoto et al.  
1998; Nakamura et al. 1999, 2000a,b).       
 
Nomoto et al. (2000) and Nakamura et al. (2000b) have shown that, while 
the highly energetic model can explain the early light curve, a less energetic 
model is in better agreement with the late light curve. They suggest 
that such a behavior reflects some non--spherical effects in the ejecta structure, 
although the well mixed models of Sollerman et al. (2000) do not require asymmetries 
for a reasonable match to the light curves, neither do those of Chugai (2000).

Jeffery's (1999) {\it ad--hoc} model, although based on a rather 
different model of the ejecta, also predicts that the pure exponential 
phase should end after about day 400, and that the light curve should 
slowly approach the $^{56}$Co decay slope when eventually only the 
positrons are deposited (provided that other, less abundant 
radioactive species remain unimportant and interaction with 
circumstellar material does not occur).  In this respect it is interesting 
to note that a least squares fit to the bolometric light curve in the phase  
range 376--490 days gives a slope $\gamma=0.009\pm0.001$ mag day$^{-1}$,  
which would suggest that SN~1998bw finally settled on the $^{56}$Co decay.  
Nevertheless a possible alternative is that the flattening is due to the 
unaccounted presence of other sources within the ground--based point spread 
function. In order to reproduce the observed behavior, the integrated magnitude of
the contaminating objects must be  
V$\sim$22.6. Even though recent HST--STIS observations (Fynbo et al. 2000)
have shown that several objects are present within a radius of 0\as5 from
the SN location (Fig.~\ref{fig:map}), there is no evidence for such a
relatively bright source. Hence the conclusion is that the observed flattening 
is, at least to some extent, intrinsic. However, this can only be confirmed by 
follow--up observations by HST. On the other hand, the effects of possible 
freeze--out, ejecta--wind interaction or even faint echoes should not be overlooked.

\begin{figure*}
\figcaption[fig17.eps]{\label{fig:lclate} BVRI broad band light curves of   
SN~1998bw at $t>$40 days. For presentation the light curves have been 
shifted  by the reported amounts. No extinction correction has been applied.  
The dotted and dashed lines represent least--squares fittings to the data in 
the ranges 40--330 and 300--490 days past $B$ maximum.  Data are from Galama 
et al. (1998b), McKenzie \& Schaefer (1999), Sollerman et al. (2000) and this  
work. The thick dashed line corresponds to the 
$^{56}\rm Co \rightarrow ^{56}$Fe decay rate, expected for full 
$\gamma$--ray trapping. Epochs refer to the $B$ maximum light.}  
\end{figure*}

As a matter of fact, no signs of ejecta--wind interaction are found in the late phase
spectra of SN~1998bw (see also Sollerman et al. 2000). Nevertheless, some circumstellar 
material is expected to be present. All models for SNe~Ic assume that the progenitor 
stars undergo intense mass-loss and lose their H--He envelopes to 
become bare CO cores before they explode. Thus interaction should be expected at some 
stage. Future observations will be fundamental to investigate the circumstellar 
environment and the real onset of ejecta-wind interaction. 
 
In Fig.~\ref{fig:abslc} the absolute V light curve of SN~1998bw is presented,   
and compared with the light curves of some Type II (1987A, 1979C), IIb  
(1993J), Ia (1991T, 1992A), Ib (1990I) and Ic (1992ar, 1994I) SNe. For  
SN~1998bw a distance modulus $\mu=32.89$   
while the interstellar extinction is assumed to be $A_V=0.2$ mag (cf. 
Sec.~\ref{sec:highres}).  
It is clear that the light curve alone would not allow one to classify this  
object as a SN~Ib/c, nor indeed to distinguish it from thermonuclear SNe.  
  
Even though the internal extinction for SN~1994I is very uncertain 
(here we adopted $A_V=1.2$), SN~1998bw was much brighter and had a different 
light curve shape. Unfortunately, late phase data for SN~1994I are not 
available. However, a comparison is possible with the Type Ib 
SN~1990I. Maximum light was not covered by the observations, but the 
light curve of SN~1990I between day +40 and day +120 is very similar 
to that of SN~1998bw, although about 0.5 magnitudes fainter. After 
that the luminosity decay rate of SN~1990I was larger than that of 
SN~1998bw. Later on the over luminosity of this SN becomes 
more pronounced. At 350 days the Type IIb 
SN~1993J is 1.4 magnitudes fainter then SN~1998bw (see Fig.~\ref{fig:abslc}). 
SN~1996N is even less luminous as shown by Sollerman et al. (1998), 
even though the distance modulus, reddening and epoch of maximum are rather 
uncertain. One year after the explosion this SN is about 3 magnitudes fainter 
than SN~1998bw.
  
As discussed before, we believe that the late time light curve of 
SN~1998bw is powered by the radioactive decay of $^{56}$Co into 
$^{56}$Fe and therefore the late luminosity can be used to constrain 
the $^{56}$Ni mass ejected by the explosion. 
 The decay of $^{56}$Co $\rightarrow$ $^{56}$Fe releases energy through the 
$\gamma$-ray channel in 81\% of the cases, while the remaining fraction goes 
into positrons, which annihilate with electrons producing 
$\gamma$-rays (cf. Colgate \& McKee 1969). The positron kinetic energy 
released before the annihilation accounts for 3.5\% of the total 
$^{56}$Co decay energy (Arnett 1979, Woosley, Pinto \& Hartmann 1989). 
  
In the case of SN~1998bw, since the slope of the late light curve  
prior to 300 days is steeper than the expected input energy from $^{56}$Co  
decay, we can assume that in this phase range some fraction of the  
$\gamma$-rays escapes thermalization. At later epochs ($t>$400 days) 
the envelope becomes completely transparent to $\gamma$-rays whereupon in  
the case of complete deposition of positron kinetic energy (Axelrod 1980),  
the late light curve settles on the $^{56}$Co decay line. 
 
Making the crude assumption that the phase of complete escape of 
$\gamma$--rays and complete deposition of positrons was reached after 400 
days and assuming a rate of $^{56}$Co decay energy production   
of $s$=6.78$\times$10$^9$ erg g$^{-1}$ s$^{-1}$ (Sutherland \& Wheeler 1984)
and a half--life time of 77.12 days (Arnett 1996),
we can estimate the mass of $ M(^{56}\rm Ni)$ using the following equation: 
 
\begin{equation}  
\label{eq:nimass}  
 M(^{56}\rm Ni) \lesssim \frac{L_{43}(t)}{1.35 \; e^{-t/111.26} 
\times 0.035} \;\; M_\odot  
\end{equation}  
   
where $L_{43}(t)$ is the bolometric luminosity in $10^{43}$ 
erg s$^{-1}$, $t$ is expressed in days from the explosion and the factor 0.035 
is the fraction of total $^{56}$Co decay energy deposited by positrons. 
If we use the four available measurements at $t>$ 400 days (Sollerman et al.  
2000; this work) and average the  
results, we get $M(^{56}\rm Ni) \leq 1.0^{+0.5}_{-0.2} \; M_\odot$. The errors  
are by far dominated by the uncertainties in the bolometric luminosities and 
the estimate depends on our assumption on the IR contribution at these  
late phases.  
 
We note that the model by Nakamura et al. (2000b) estimated the  
mass of $^{56}$Ni to be 0.4 M$_\odot$ from the early light curve modeling.  
This value is not in contradiction with what is found here if the SN envelope  
is not completely transparent to $\gamma$--rays at $t>$ 400 days. Moreover,
it must be noted that Nakamura et al. (2000b) have suggested that positron
contribution is not yet dominant in powering the light curve at the
phases covered by the last available observations.

\begin{figure*}
\figcaption[fig18.eps]{\label{fig:lcmodel} Comparison between the UVOIR  
bolometric light  curve of SN~1998bw and the hypernova model by Iwamoto  
et al. (1998, solid line). The dotted line represents an extrapolation of a  
least--squares fit to the data in the range 50--200 days while the  
dotted--dashed line is a fit to the data in the phase range 376--490 days. 
The dashed line corresponds to the $^{56}\rm Co \rightarrow ^{56}$Fe decay 
rate, expected for full $\gamma$--ray trapping. The plot in the upper right 
part of the figure shows the deviations of the observed data from the 
extrapolation on the early light curve (50--200 days, see text). 
For comparison the bolometric light curve of SN~1987A (Bouchet et al. 1991)  
is also plotted (filled triangles).}  
\end{figure*}  
 
\begin{figure*}
\figcaption[fig19.eps]{\label{fig:abslc} Absolute V light curve of SN~1998bw   
compared with SNe 1987A, 1979C (Balinskaya et al. 1980, De Vaucouleurs et al. 
  1981, Barbon et al. 1982), 1993J (IAU Circulars; Lewis et al. 1994; Barbon 
et al. 1995), 1991T (Phillips et al. 1992, Cappellaro et al.   
1999), 1992A (Suntzeff 1996, Cappellaro et al. 1999), 1990I (ESO-KP  
data base, unpublished), 1992ar (Clocchiatti et al. 2000a), 1994I  
(Tsvetkov \& Pavlyuk 1995, Richmond et al. 1996).   
Note that for SN~1992ar the brightest case was chosen (cf. Clocchiatti  
et al. 2000). 
The dotted lines represent a least--squares fit to the late phase data of   
SNe 1991T and 1992A. The dashed line corresponds to the $^{56}\rm Co   
\rightarrow ^{56}$Fe decay rate, expected for full $\gamma$--ray trapping.}  
\end{figure*}

\section{\label{sec:disc} Discussion and Conclusions}  
  
As we have shown, SN~1998bw was exceptional in many respects, even 
beyond its possible and probable connection 
with GRB~980425. The luminosity at maximum was comparable to that of 
a SN~Ia but its spectral appearance was completely different from that 
class of objects. On the other hand, while the Type Ic classification at 
maximum holds by definition (no H or He, weak Si~II), SN~1998bw had 
little in common with objects previously classified as Type Ib or Ic 
(cf. also Stathakis et al. 2000), at least around maximum. Among all studied 
Type Ib/c SNe,  only SN~1992ar might have been brighter than SN~1998bw 
(Clocchiatti et al. 2000a). The only known SN which bear some spectroscopic 
resemblance to this unusual object are SN~1997ef (Garnavich et al. 1997) and 
SN~1998ey (Garnavich, Jha \& Kirshner 1998), the former also being possibly 
associated  with a GRB (Wang et al. 1998). Even so, SN~1997ef was fainter 
than SN~1998bw, and probably produced much less $^{56}$Ni than SN~1998bw 
(Iwamoto et al. 1998, 2000). It must be mentioned here that SN~1997cy, 
also conceivably but by no means certainly associated with a GRB 
(Woosley et al. 1999, Germany et al. 2000), 
was extremely bright, probably the brightest SN ever observed, but it has 
shown signs of ejecta--CSM interaction at all epochs (Turatto et al. 2000). 
 
The deviation from any known Type Ib SN behavior is also noticeable 
in the early nebular phase, when the spectrum of SN~1998bw is 
probably dominated by Fe blends in the blue and by O and Ca in the red, 
as already pointed out by Patat and Piemonte (1998b). A somewhat similar 
behavior was noticed in SN~1993R by Ruiz-Lapuente (see Filippenko 1997a) 
and in SN~1990aj by Piemonte (2001). 
  
The high velocities and the large intrinsic luminosity suggest that SN~1998bw was 
produced by an extremely energetic explosion. All published models reach this 
conclusion, even if the explosion energy, the progenitor mass and ejected 
$^{56}\rm Ni$ mass, span 
quite a wide range (Iwamoto et al. 1998, H\"oflich et al. 1999, 
Woosley et al. 1999, Nakamura et al. 2000b).  This range arises because 
different degrees of asymmetry and beaming have been used. 
The exceptionally high value of the kinetic energy 
of the models which give the best fit to both the light curve and the 
spectra of SN~1998bw ($3.0\times10^{52}$ erg, Iwamoto et al. 1998; 
$5.7\times10^{52}$ erg, Nakamura et al. 2000a,b; $6.0\times10^{52}$ erg, 
Branch 2001) led to the designation of this object as a hypernova,
even though it does not match the original hypernova definition (Paczy\'nski 1998).

Despite the great individuality shown by SN~1998bw at maximum light, 
one year later its spectrum is very similar to that of other Type Ib/c 
events.  This gives qualitative support to the idea that the material 
ejected by SN~1998bw was rich in both Fe--peak and $\alpha-$elements. 
The possibly large production of Fe-peak elements makes it  
important to obtain detailed nucleosynthesis calculations and rate 
estimates of occurrence of such objects, because they 
might have a significant impact on models of galactic chemical 
evolution.  
Since SN~1998bw was as bright as a SN~Ia, it might seem unlikely that 
such objects would be missed in nearby searches. Nevertheless 
SN~1998bw may have been missed had there not been an associated 
GRB. Does this allow us to infer that such kind of explosions must be 
intrinsically rare at the present epoch? 
  
But even if they are relatively rare now, their past frequency may 
have been considerably higher. SN~1998bw is thought to have been 
generated by a massive star (cf. Iwamoto et al. 1998, Woosley et al. 1999)  
which may have been more common at remote epochs, when the rate of star  
formation was higher. Since the time scale for the release of iron and oxygen 
into the ISM by these objects is much shorter than 
for SNe~Ia and significantly less massive core-collapse SNe, 
hypernovae may have played an important role in the initial galactic 
chemical enrichment.  Future work will concentrate on establishing whether the rate 
of occurrence of such SNe increases with redshift or look--back time. It is 
possible that the discovery of many GRB's at significant redshift is already a 
hint in this direction, as is the conclusion for several of them that the light
curve can be reproduced with a combination of a GRB afterglow plus a SN
light curve (cf. Bloom et al. 1999).
  
\acknowledgements{  
  
This work is based on data collected at ESO-La Silla. We express our 
gratitude to the Visiting Astronomers who kindly gave us part of their 
observing time in order to secure a good follow-up of this important 
object. In particular, we acknowledge the support we received from 
Pierre Leisy and Alessandro Pizzella during the observations at La Silla. 
We also wish to thank Stephen Holland and Jens Hjorth for making the 
HST--STIS image of SN~1998bw available to us. 
Finally, the authors gratefully acknowledge an anonymous referee for 
the thorough review, which really helped to improve the paper.

This work has been partially supported by the grant-in-Aid for Scientific  
Research (07CE2002, 12640233) of the Ministry of Education, Science and  
Culture and Sports in Japan. The authors made use of the NASA/IPAC  
Extragalactic Database (NED) which is operated by the Jet Propulsion  
Laboratory, California Institute of Technology, under contract with the 
National Aeronautics and Space Administration.}

\clearpage

\bibliographystyle{mn}

\clearpage

    
\begin{table*}  
\caption{\label{tab:spec}Journal of spectroscopic observations.}  
\bigskip  
\scriptsize  
\begin{center}  
\begin{tabular}{ccccccc}  
\tableline  
\tableline  
Date     & JD         & Phase\tablenotemark{a}  & Range     &   
Resolution\tablenotemark{b} & Equipment & Standard Stars \\  
         & (2400000+) &(days)     & (\AA)      & FWHM (\AA)     &                 & \\  
\tableline  
1.3 May 98   & 50934.8 & $-$9       & 5900-9200  & 10             & NTT-EMMI      & G138-31\\  
3.4 May 98   & 50936.9 & $-$7       & 3350-9000  & 10             & Danish        & LTT7379\\  
4.3 May 98   & 50937.8 & $-$6       & 3400-10250 & 20             & ESO-3.6m      & CD32d9927\\  
4.4 May 98   & 50937.9 & $-$6       & 3400-7550  & 20             & ESO-3.6m-Pol. & HD161056\\  
6.3 May 98   & 50939.8 & $-$4       & 3750-7650  &  1             & NTT-ECH       & -\\  
7.3 May 98   & 50940.8 & $-$3       & 3500-9000  & 10             & Danish        & LTT7379\\  
8.3 May 98   & 50941.8 & $-$2       & 3350-10150 & 10             & Danish        & LTT7379\\  
9.3 May 98   & 50942.8 & $-$1       & 3300-10150 & 10             & Danish        & LTT7379\\  
11.2 May 98  & 50944.7 & +1         & 3450-10200 & 10             & Danish        & LTT7379\\  
13.4 May 98  & 50946.9 & +3         & 3450-10200 & 10             & Danish        & LTT7379\\  
14.4 May 98  & 50947.9 & +4         & 3350-8900  & 10             & Danish        & LTT7379\\  
16.4 May 98  & 50949.9 & +6         & 3400-9000  & 10             & Danish        & LTT7379\\  
18.3 May 98  & 50951.8 & +8         & 9500-25200 & 18             & NTT-SofI      & Hip106725\\  
19.4 May 98  & 50952.9 & +9         & 3250-9000  & 10             & Danish        & EG274\\  
20.3 May 98  & 50953.8 & +10        & 3400-7550  & 20             & ESO-3.6m-Pol. & HD161056\\  
21.4 May 98  & 50954.9 & +11        & 3400-9000  & 10             & Danish        & LTT6248\\  
22.4 May 98  & 50955.9 & +12        & 3400-9000  & 10             & Danish        & LTT7379\\  
23.3 May 98  & 50956.8 & +13        & 3100-10200 & 15             & ESO-1.52      & Feige110\\  
29.4 May 98  & 50962.9 & +19        & 3350-10250 & 20             & ESO-3.6m      & LTT7379\\  
1.3 Jun. 98  & 50965.8 & +22        & 3400-9000  & 10             & Danish        & LTT6248\\  
8.2 Jun. 98  & 50972.7 & +29        & 4200-9000  & 10             & Danish        & LTT6248\\  
12.3 Jun. 98 & 50976.8 & +33        & 9500-25200 & 18             & NTT-SofI      & G2720\\  
24.2 Jun. 98 & 50988.7 & +45        & 3400-9000  & 10             & Danish        & LTT6248\\  
30.3 Jun. 98 & 50994.8 & +51        & 9500-25200 & 18             & NTT-SofI      & BS4620,BS6823\\  
1.2 Jul. 98  & 50995.7 & +52        & 3400-9000  & 10             & Danish        & LTT7987\\  
13.2 Jul. 98 & 51007.7 & +64        & 3400-9000  & 10             & Danish        & LTT6248\\  
22.3 Jul. 98 & 51016.8 & +73        & 3100-10100 & 15             & ESO-1.52      & Feige110\\  
12.2 Aug. 98 & 51037.7 & +94        & 3100-10100 & 15             & ESO-1.52      & Feige110\\  
12.2 Sep. 98 & 51068.7 & +125       & 3400-10200 & 20             & ESO-3.6m      & Feige110\\  
26.0 Nov. 98 & 51144.5 & +201       & 3350-10200 & 20             & ESO-3.6m      & LTT1020\\  
12.4 Apr. 99 & 51280.9 & +337       & 3350-10250 & 20             & ESO-3.6m      & LTT6248,LTT7379\\  
21.2 May 99  & 51319.7 & +376       & 3350-10250 & 17             & ESO-3.6m      & LTT3864,Feige110\\  
21.2 May 99  & 51319.7 & +376       & 4750-6750  &  7             & ESO-3.6m      & LTT3864,Feige110\\  
\end{tabular}  
\end{center}  
\tablenotetext{a}{\scriptsize Relative to B maximum (JD=2450943.8)}   
\tablenotetext{b}{\scriptsize FWHM of night sky lines}  
\tablecomments{\scriptsize  
NTT-EMMI = ESO-NTT+EMMI+TK2048 (red arm) +TH1024 (blue arm),   
NTT-ECH = NTT + EMMI echelle mode + TK2048,  
Danish =  ESO-Danish 1.54 + DFOSC + Loral/Lesser2048,  
ESO-3.6m = ESO-3.6m + EFOSC2 + Loral/Lesser2048,  
ESO-3.6m-Pol. = ESO-3.6m + EFOSC2 + Polarimeter + Loral/Lesser2048,  
NTT-SofI = ESO-NTT + SofI + Rockwell Hg:Cd:Te1024,  
ESO-1.52 = ESO 1.52m + B\&C + Ford2048}  
\end{table*}

\begin{table*}  
\small  
\caption{\label{tab:irphot} IR photometry of SN~1998bw.}  
\bigskip 
\centerline{ 
\begin{tabular}{cccccccccc}  
\tableline  
\tableline  
Date     &JD        &Phase\tablenotemark{a} &   
{\it J}     &  {\it H}    &  {\it K}    & Seeing   & Telescope\\  
         &(2400000+)&(days)    &       &       &       & (arcsec) & \\  
\tableline  
18.3 May  98 & 50951.8  &+8        & 13.40$\pm$0.04 & 13.35$\pm$0.04 & 13.15$\pm$0.04 & 0.7 & NTT+SofI\\  
12.3 Jun. 98 & 50976.8  &+33       & 14.41$\pm$0.04 & 14.25$\pm$0.04 & 14.20$\pm$0.04 & 1.2 & NTT+SofI\\  
30.3 Jun. 98 & 50994.8  &+51       & 15.11$\pm$0.04 & 14.98$\pm$0.04 & 14.89$\pm$0.04 & 0.9 & NTT+SofI\\  
\end{tabular}  
}
\tablenotetext{a}{Relative to B maximum (JD=2450943.8). This occurred  
14.4 days after GRB~980425.}  
\end{table*}  
  
  \begin{table*}  
\caption{\label{tab:photo} Late Phase Photometric observations of SN~1998bw.}  
\bigskip  
\scriptsize  
\centerline{
\begin{tabular}{cccccccccc}  
\tableline  
\tableline  
Date     &JD        &Phase\tablenotemark{a}&   
{\it U}  &  {\it B}  &  {\it V}  &  {\it R}  &  {\it I}  & Seeing & Telescope \\  
         &(2400000+)&(days)   &    &     &     &     &     & (arcsec) & \\  
\tableline  
16.1 Mar. 99 &51253.6&+310& - & 20.69$\pm$0.07 & 20.52$\pm$0.07 & - & - & 1.3 & Dutch-0.9\\  
17.1 Mar. 99 &51254.6&+311& - & 20.71$\pm$0.07 & 20.50$\pm$0.07 & 19.74$\pm$0.05 & - & 1.3 & Dutch-0.9\\  
8.1 Apr. 99 &51276.6&+333& - & 21.10$\pm$0.15 & 20.69$\pm$0.15 & 20.09$\pm$0.15 & - & 1.1 & ESO-3.6\\  
12.2 Apr. 99 &51280.7&+337& 21.13$\pm$0.20 & - & - & - & 20.03$\pm$0.15 & 1.1 & ESO-3.6 \\  
21.2 May 99 &51319.7&+376& - & 21.59$\pm$0.20 & 21.43$\pm$0.20 & 20.83$\pm$0.20 & 20.61$\pm$0.20 & 1.2 & ESO-3.6 \\  
17.3 Jun. 99 &51346.8&+403&21.69$\pm$0.20 & 21.91$\pm$0.20 & 21.70$\pm$0.20 & 20.87$\pm$0.20 &  20.76$\pm$0.20 & 1.0 & ESO-3.6 \\  
\end{tabular}  
}
\tablenotetext{a}{\scriptsize Relative to B maximum (JD=2450943.8).   
This occurred 14.4 days after GRB~980425.}  
\tablecomments{\scriptsize Dutch-0.9=0.92m ESO-Dutch + CCD TK512,   
ESO-3.6=ESO-3.6m + EFOSC2 + Loral/Lesser2048}  
\end{table*}

\begin{table}
\caption{\label{tab:slopes} Decline rates of SN~1998bw in the phase ranges 40--330 
and 300--490 days form B maximum.}
\bigskip
\centerline{
\begin{tabular}{lcccc}
\tableline
\tableline
            &   B    &    V     &    R    &   I \\
\tableline
Phase Range\tablenotemark{a} & 40.8-311.6        & 40.8-325.6        &  40.8-325.6        & 40.8-325.6 \\
N. of points                 & 51                & 57                &  7                 & 47 \\
$\gamma$  (mag day$^{-1}$)   & 0.0150$\pm$0.0010 & 0.0180$\pm$0.0006 &  0.0165$\pm$0.0019 & 0.0169$\pm$0.0010\\
            &        &          &         &      \\
Phase Range\tablenotemark    & 310.6-402.6       & 310.6-489.6       & 311.6-489.6        & 325.6-489.6\\
N. of points                 & 5                 & 10                &  9                 & 8 \\
$\gamma$  (mag day$^{-1}$)   & 0.0130$\pm$0.0008 & 0.0101$\pm$0.0008 & 0.0129$\pm$0.0006  & 0.0126$\pm$0.0008\\
\end{tabular}
}
\tablenotetext{a}{Relative to B maximum (JD=2450943.8).}  
\end{table}

\end{document}